\documentclass[prl,twocolumn,showpacs,floatfix,amssymb]{revtex4}
\usepackage{graphics}
\begin{document}

\title{\hfill\underline{{\small submitted to Phys.Rev.Lett.}}\\
Evidence of strong electron-phonon interaction in superconducting MgB$_2$ from electron
tunneling.}

\author{A. I. D'yachenko}
\altaffiliation[permanent address: ] {Donetsk Physico-Technical Institute, Ukrainian
National Academy of Sciences, R.Luxemburg St. 72, 340114 Donetsk, Ukraine.}
\author{V. Yu. Tarenkov}
\altaffiliation[permanent address: ] {Donetsk Physico-Technical Institute, Ukrainian
National Academy of Sciences, R.Luxemburg St. 72, 340114 Donetsk, Ukraine.}
\author{A. V. Abal'oshev} \author{S. J. Lewandowski}
\affiliation{Instytut Fizyki Polskiej Akademii Nauk, Al. Lotnik\'{o}w 32, 02-668 Warszawa,
Poland}

\begin{abstract}
We report tunneling measurements of the electron-phonon (e-ph) interaction in
superconducting MgB$_2$ using the MgB$_2$-I-Nb junctions, where I stands for insulator.
The phonon structure in tunneling density of states in MgB$_2$ clearly indicates strong
e-ph coupling for the E$_{2g}$ in-plane boron phonons in a narrow range around $60$ meV.
The Eliashberg spectral function $\alpha^2(\omega)\mathrm{F}(\omega)$ reconstructed from
the tunneling data, exhibits significant additional contribution into e-ph interaction
from other vibrations such as acoustic ($\sim 38$ meV) and optical ($\sim 90$ meV) bands.
Our results are in reasonable agreement with neutron scattering experiments, and also to
some data of Raman and infrared spectroscopy.

\end{abstract}
\maketitle

The recent discovery of superconductivity near $40$K in MgB$_2$, stimulated a renewed
interest in the pairing mechanism of cuprate superconductors. It is generally believed
that understanding superconductivity in the simple binary MgB$_2$ compound should be much
easier than in the high-T$_c$ superconductors. Theoretical {\it ab initio} calculations
show that dominant contribution to Eliashberg function in MgB$_2$ arises from the
Raman-active E$_{2g}$ phonons near the Brillouin zone center at \ $\sim 75$~meV
\cite{Kong, An, Bohnen, Liu, Kortus}. This conclusion is supported by Raman studies
\cite{Goncharov, Bohnen, Hlinka, Kunc}, which indicate a mode at \
$\sim600-620~\mathrm{cm^{-1}}$ ($74 -77$~meV) with a very large width \ $\sim
200~\mathrm{cm^{-1}}$ ($25$~meV), usually related to anharmonic effects. However, there
has been no consensus about the energy and role of the E$_{2g}$ mode, because this mode
has not been identified unambiguously by inelastic neutron scattering or in Raman
experiments \cite{Chen, Kunc, Rafailov, Struzhkin, Meletov}.

Tunneling spectroscopy can provide a direct measure of the Eliashberg electron-phonon
spectral function $\alpha^2(\omega)\mathrm{F}(\omega)$, which can reveal the importance
of different phonon modes in the superconductivity of MgB$_2$. Here F$(\omega)$ is phonon
density of states and $\alpha(\omega)$  is an effective electron-phonon coupling function
for phonons of energy $\hbar\omega$. Tunneling conductivity $\sigma =
\mathrm{d}I/\mathrm{d}V$ gives direct information about the density of states
$N(\omega)\propto\mathrm{Re}\{\omega/[\omega^2- \Delta(\omega)^2]^{1/2}\}$ related to
quasi-particle excitations in the superconductor, where $\Delta(\omega)$ is a complex and
energy dependent gap function \cite{Schrieffer}. $\alpha^2(\omega)\mathrm{F}(\omega)$
 can be obtained by the inversion of the Eliashberg equations for
$\Delta(\omega)$ \cite{Mcmillan,Galkin}. The method is based on the determination of
$\Delta(\omega)$, therefore it non-sensitive to the presence of non-superconducting
impurity phases in MgB$_2$. The latter can give comparable features in Raman and infrared
spectra at energy range typical for the superconducting MgB$_2$ \cite{Rafailov, Sundar}.

We report here high resolution superconducting tunneling spectroscopy measurements of the
phonon structure of MgB$_2$. We used tunnel point-contact junctions Nb-I-MgB$_2$, where I
is a natural tunnel barrier. We found that the second derivative of tunnel current
d$^2I/$d$V^2$ exhibits features, which coincide with peaks in the phonon density of
states (DOS) F$(\omega)$ obtained from neutron scattering \cite{Osborn, Yildirim,
Clementyev}, Raman \cite{Meletov} and infrared \cite{Tu, Sundar} spectroscopy. Numerical
treatment of the experimental data indicates that the observed phonon structures between
$55-65$~meV and near $90$~meV are very strongly coupled to the electrons.

\begin{figure} \includegraphics{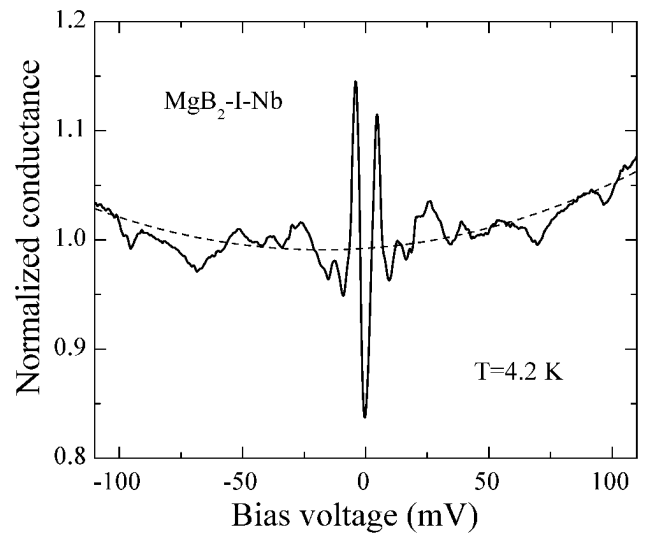}
\caption{Experimental conductance of MgB$_2$-I-Nb tunnel junction in superconducting
state (solid line) and the corresponding approximation of the normal state data (dashed
line).} \label{fig1}
\end{figure}

 The investigated samples were prepared by compacting commercially
 available MgB$_2$ powder  (Alfa Aesar, purity $98\%$) at high pressure ($P>5$~GPa)
 into thin rectangular bars of about
$1\times0.6\times0.08 \ \mathrm{mm}^3$. Details of the sample preparation are described
elsewhere \cite{Dyachenko}. The samples revealed a superconductive transition temperature
(onset) at T$_c=38$~K and a high critical current density $J_c>10^5 \mathrm{A/cm^2}$ at
T$ \ =4.2$~K, considerably larger than the current densities employed in the tunneling
measurements. Therefore, current depairing had no effect on the phonon-mediated region of
tunneling data. The point-contact junctions were obtained by pressing Nb wire
($\varnothing \sim 50 \mu$m) into the surface of the sample. The tunnel characteristics
were measured using a standard four probe lock-in technique.

In a number of Andreev S-N-S and  tunnel S-I-S experiments at T$ \ =4.2$~K we observed
two gaps $\Delta$, the smaller gap value of about $3$~meV, and a larger gap close to
$7$~meV, both results in agreement other measurements \cite{Buzea}. In tunnel S-I-S
junctions we observed usually only the smaller gap with s-wave symmetry. Let us observe
that the smaller $\sim 3$ meV and the larger $\sim 7$ meV gaps are associated with the
$3d$ parts and quasi-2D sheets of the Fermi surface, respectively \cite{Liu}. The
experimental detection of the single small gap is possible in the case of tunneling
perpendicular to the honeycomb born planes. Due to the strong directionality of the
tunnelling experiments, the probability of normal injection is maximal and electronic
properties are mainly probed along the direction perpendicular to the junction surface,
i.e. in our case normal to the born planes.

At high bias voltages ($> 20$ meV) the d$I/$d$V(V)$ data reveal a structure (Fig.1),
which in the case of conventional superconductors is characteristic of phonon effects
\cite{Wolf}. To investigate this structure in more detail, we used the second derivative
of the tunneling current d$^2I/$d$V^2$, obtained by standard harmonic detection, as well
as by straightforward numerical differentiation of the tunneling conductivity d$I/$d$V$.
Both methods give analogous results. In these spectra, we observed sometimes peaks at
$\sim 17$ meV and $\sim 28$ meV, which are not intrinsic to clean MgB$_2$. The structure
near 28 meV most likely arises from a thin pure Mg layer in proximity with the
superconducting MgB$_2$ (analogous structures have been observed in Ag-I-Mg/Nb junctions
\cite{Burnell}). For this reason, we excluded these singularities from further
considerations, especially as they do not affect the subsequent treatment of the raw
data.

\begin{figure} \includegraphics{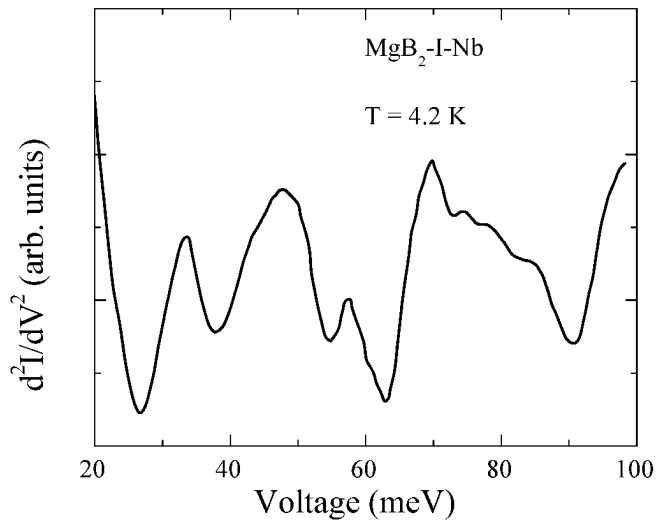}
\caption{Second derivative d$^2I/$d$V^2$ plots of a MgB$_2$-I-Nb junction. The minima in
the curve correspond to peaks in Eliashberg function
$\alpha^2(\omega)\mathrm{F}(\omega)$. Voltage origin is at $\Delta_{MgB_2} +
\Delta_{Nb}$} \label{fig2}
\end{figure}

The plot d$^2I/$d$V^2(V)$, shown in Fig.~$2$, at eV~$ > 30 $~meV directly reflects the
electron-phonon spectral function $\alpha^2(\omega)\mathrm{F}(\omega)$ of MgB$_2$; in
particular the dips in d$^2I/$d$V^2$ correspond to peaks in
$\alpha^2(\omega)\mathrm{F}(\omega)$. Peaks in F$(\omega)$ should be reproduced in
$\alpha^2(\omega)\mathrm{F}(\omega)$ at approximately the same energy, even if the
coupling factor $\alpha^2(\omega)$ is strongly energy dependent. The d$I/$d$V$ and
d$^2I/$d$V^2$ data were used to obtain the normalized tunneling conductance
$\sigma/\sigma_N$, where $\sigma_N$ is a normal-state (background) conductance, and the
tunneling density of states $N(\omega)$; the relevant numerical method is given in
\cite{Wolf}.

\begin{figure} \includegraphics{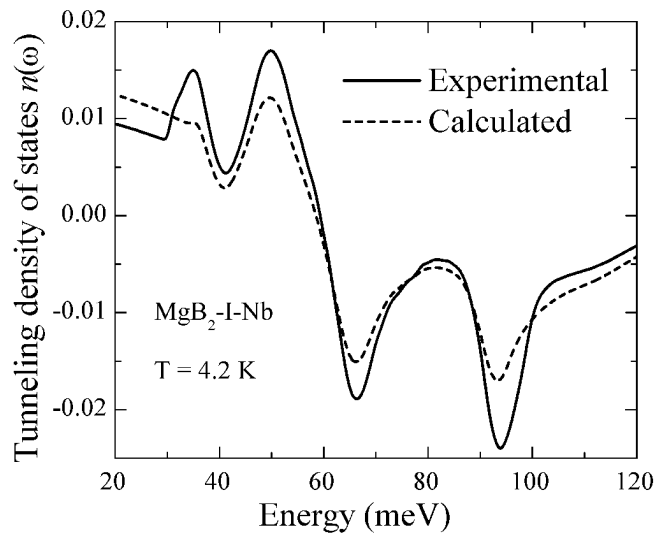}
\caption{The tunneling density of states of MgB$_2$ $n(\omega)$, normalized to the BCS
density of states. Experimental data are plotted as a solid line; dashed line shows
$n(\omega)$ calculated from Eliashberg function $\alpha^2(\omega)\mathrm{F}(\omega)$.}
\label{fig3}
\end{figure}

The normalized density of states $n(\omega) = N(\omega)/N_{\mathrm{BCS}}(\omega) - 1$,
proportional to the smoothed d$I/$d$V$ curve, is shown in Fig. 3. Here
$N_{\mathrm{BCS}}(\omega)$ is the broadened BCS density of states (Dynes function). The
d$I/$d$V$ curve displays a well-known behavior due to the phonon mode reflection in
tunneling S-I-S conductance \cite{Mcmillan,Schrieffer}, namely, sharp drops of d$I/$d$V$
near the phonon emission thresholds, to which correspond dips in d$^2I/$d$V^2$. We can
rule out very weak inelastic process of phonon-assisted tunneling, because such process
increases the tunneling conductance near the peaks in F$(\omega )$ and gives
corresponding peaks in d$^2I/$d$V^2$ \cite{Wolf}. The common feature of all measured
d$^2I/$d$V^2$ curves is the reproducibility of dip positions at $\sim 40$, $55\div65$ and
$90\div92$ meV, measured from the gap sum $\Delta_{MgB_2} + \Delta_{Nb}$ (Fig.2). There
is a close correlation in the peak positions in our d$^2I/$d$V^2$ data and thouse in the
generalized phonon density of states (GPDOS) as measured by inelastic neutron scattering
data \cite{Osborn, Yildirim}. The first tunneling peak near $40$ meV arises from acoustic
modes of MgB$_2$. The well-resolved tunneling structures at $\sim 60$~meV and $\sim
90$~meV are associated with optical vibration of the born atoms. As it is well known, in
the vicinity of peaks in $\alpha^2(\omega)\mathrm{F}(\omega)$, the gap function
$\Delta(\omega)$ has a strong energy dependence and large Im$\{\Delta(\omega)\}$
\cite{Schrieffer}. At  $\hbar\omega \gg |\Delta|$  tunneling density $N(\omega) \sim
1+(1/2)\mathrm{Re}\{\Delta^2(\omega)/\omega^2\}$, so the phonon structure in $N(\omega)$
is weighted by the factor $(\delta\Delta/\omega)^2$. In the strong coupling limit (the
McMillan's electron-phonon coupling constant $\lambda$  of the order of $1$) variation of
$\Delta(\omega)$, $\delta\Delta \sim \Delta$, and the deviation in $N(\omega)$ is of the
order $1-2\%$ (Fig.3). Even without detailed calculations one can see that since the
amplitude of tunneling structure at $40$ meV is roughly one half of that at $60$ meV
(c.f. Fig.2) and the optical phonon energy is $1.5$ times higher, the actual electronic
coupling to the phonons at $60$ meV may be as much as $(1.5)^2\sim 3$ times greater than
that for acoustic phonons at $\sim 40$~meV. On the other hand, it is evident, that the
optical modes at $\sim 90$~meV give approximately the same contribution than that of
modes at $60$ meV. Such conclusion, obtained directly from the tunneling data, is
independent of the anisotropy of the energy gap $\Delta$({\bf k}) and agrees well with
similar estimations based upon Raman and infrared spectra measurements \cite{Tu, Sundar}.

\begin{figure} \includegraphics{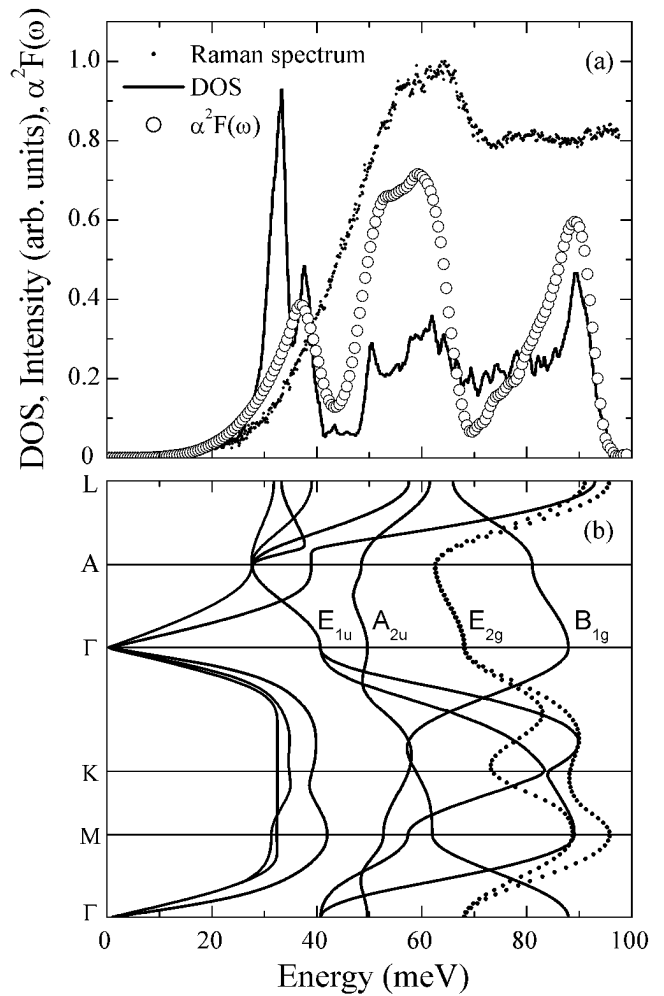}
\caption{Comparison of the phonon density of states in MgB$_2$ obtained by different
methods. (a) The function $\alpha^2(\omega)\mathrm{F}(\omega)$ of MgB$_2$, determined
from tunneling measurements in this work (open circles). The continues curve is the
phonon density of states F$_t(\omega)$ calculated from neutron data \cite{Yildirim}. The
dotted curve shows Raman spectra of MgB$_2$ at room temperature \cite{Meletov}, for the
sample obtained under high pressure. (b) The calculated phonon dispersion curves
\cite{Yildirim}.} \label{fig4}
\end{figure}

An exact determination of $\alpha^2(\omega)\mathrm{F}(\omega)$ requires full inversion of
the Eliashberg equations. We used normalized conductivity $\sigma/\sigma_N$ for
self-consistent calculation of the spectral function $\alpha^2(\omega)\mathrm{F}(\omega)$
with a standard inversion program \cite{Wolf}, which allows to take into account the
already mentioned possible presence of a thin Mg layer on the surface of MgB$_2$, in
which reduced superconductivity is induced by proximity effect. The gap edge region below
$26$ meV was cut off to avoid a divergence of DOS. The obtained "experimental" function
$\alpha^2(\omega)\mathrm{F}(\omega)$ is shown in Fig.4 together with theoretical {\it ab
initio} phonon density of states F$_t(\omega )$ combined with neutron scattering
\cite{Yildirim}. In isotropic limit for $\lambda = 0.9$ and $\mu^* = 0.1$ Eliashberg
function yields T$_c \approx 38$~K, and $|\Delta| = 6.5$~meV. Neutron GPDOS \cite{Osborn,
Yildirim} differs in its detailed shape from F$_t(\omega)$, but peaks in F$_t(\omega)$,
GPDOS and our $\alpha^2(\omega)\mathrm{F}(\omega)$ all occur at the same energies. They
all show pronounced minima at $40\div50$ meV and maxima at $37$, $62$ and $90\div92$~meV.
The peak in $\alpha^2(\omega)\mathrm{F}(\omega)$, responsible for acoustic bands ($\sim
38$~meV), is strongly suppressed, but does not disappear completely, as in the
theoretical calculations \cite{Liu, Kong}. We do not see also the prevailing amplitude of
one mode at $60-70$~meV, as predicted for the theoretical function
$\alpha^2(\omega)\mathrm{F}_t(\omega)$ \cite{Liu,Kortus,Kong}, although our experimental
data indicate two bands of optical modes corresponding primarily to the born motions at
$\sim 60$ and $\sim 90$~meV. For comparison, in Fig.4(b)  the calculated phonon
dispersion curves are shown along the high-symmetry directions of the Brillouin zone
\cite{Yildirim}. As seen, the E$_{1u}$ and A$_{2u}$ infrared active modes give noticeable
contribution to the $\alpha^2(\omega)\mathrm{F}(\omega)$ at $40$ and $50$ meV,
respectively. They have been also observed in infrared absorption experiments at $333$
cm$^{-1}$ ($41$~meV) and $387$ cm$^{-1}$ ($48$~meV)\cite{Sundar}. Optical studies in the
c-axis oriented superconducting MgB$_2$ films show two strong phonon peaks at $380$
cm$^{-1}$ ($48$~meV) and $480$ cm$^{-1}$ ($60$~meV) \cite{Tu}. The peak in the
"experimental" $\alpha^2(\omega)\mathrm{F}(\omega)$ at $60\div65$ meV arises mainly from
the E$_{2g}$ phonon modes with wave vector along the $\Gamma - $A direction [Fig.4(b)].
This phonon mode is Raman-active. In most investigations the Raman scattering in MgB$_2$
revealed a broad peak near $620$ cm$^{-1}$ ($78$~meV) but, Meletov {\it et al}
\cite{Meletov} have observed a peak near $\sim 590$ cm$^{-1}$ ($73$~meV). They
investigated ceramic MgB$_2$ samples using a micro-Raman system, which rendered possible
the identification of small crystalline grains of MgB$_2$, whose Raman spectra differ
drastically from those of MgO or metallic Mg inclusions. After homogenisation of ceramic
samples induced by high pressure ($>5$~GPa) the E$_{2g}$ phonon mode was observed at
$500-550$ cm$^{-1}$ ($62-68$~meV) [the dotted curve at Fig. 4(a)], which is close to our
$\alpha^2(\omega)\mathrm{F}(\omega)$ result. For the E$_{2g}$ mode, there are also large
discrepancies between the results of various calculations. For example, the harmonic
energy $\omega_H(\mathrm{E}_{2g}) = 60$ meV \cite{Yildirim}, but anharmonicity leads to
an increase of $\omega_H$(E$_{2g}$) up to $75$ meV \cite{Yildirim,Kong}. Approximately
the latter value ($75$ meV) was reported in the Raman study \cite{Hlinka}, where E$_{2g}$
mode had very large ($\sim 25$~meV) width. On the other hand, our experimental
$\alpha^2(\omega)\mathrm{F}(\omega)$ coincides best with the harmonic value
$\omega_H(\mathrm{E}_{2g}) \sim 60$ meV \cite{Yildirim} (Fig.4) and is reasonably narrow.
The extremely broad E$_{2g}$ line at $\sim 620$~cm$^{-1}$ ($78$~meV) can include a sum of
peaks from impurity phases, like Mg, MgO (phonon mode of MgO at $\sim 80$~meV has been
observed in Ref. \cite{Adler}), B$_2$O$_3$, and second-order Raman signal from the
acoustic $\sim 300$ cm$^{-1}$ ($37$~meV) phonon \cite{Rafailov}. On the other hand, the
infrared absorption spectrum of MgB$_2$ is characterized by a broad band centered at
$485$ cm$^{-1}$ ($60$~meV) \cite{Sundar}, which can be associated with the peak in phonon
density of states F$(\omega)$ at $60$ meV, in agreement with our results (Fig.4).

 In summary, we have obtained high resolution
tunneling spectra of MgB$_2$ with T$_c = 38$~K. The tunneling density of states deviated
from the standard Dynes function above $30$ meV in a way, which is characteristic of
phonon effects. The second derivative curve displays dips at energies, which correspond
to peaks in the phonon density of states measured by neutron scattering an also to some
data of Raman and infrared spectroscopy. The spectral function
$\alpha^2(\omega)\mathrm{F}(\omega)$ has been reconstructed from tunneling data. The
relatively big structure in tunneling spectra confirms strong coupling of electrons to
optical E$_{2g}$ phonon modes at $\sim 60$ meV. However, other vibrations such as
acoustic ($\sim 38$ meV) and optical ($\sim 90$ meV) bands are also important and give
noticeable peaks in the tunneling spectra. Therefore, the predominance of the E$_{2g}$
phonon mode near the $\Gamma$ point in the electron-phonon mechanism of superconductivity
in MgB$_2$ may be questioned.

\begin{acknowledgments}
This work was supported by Polish Government (KBN) Grant No PBZ-KBN-013/T08/19.
\end{acknowledgments}

\end{document}